\documentclass[conference]{IEEEtran}

\usepackage{amsmath}
\usepackage{amsfonts}
\usepackage{graphicx}
\usepackage[short]{optidef}
\usepackage[margin=0.70in]{geometry}
\usepackage{cite}
\newcommand{\Fourier}{\mbox{\boldmath$\cal F$}}

\IEEEoverridecommandlockouts
\begin{document}

\title{Multi-user Capacity of Cyclic Prefix Direct Sequence Spread Spectrum with Linear Detection and Precoding}
\date{13 January 2020}
 \author{%
    \IEEEauthorblockN{Brent A. Kenney\IEEEauthorrefmark{1}, Arslan J. Majid\IEEEauthorrefmark{2}, Hussein Moradi\IEEEauthorrefmark{2}, and Behrouz Farhang-Boroujeny\IEEEauthorrefmark{1}}
    \IEEEauthorblockA{\IEEEauthorrefmark{1}Electrical and Computer Engineering Department, University of Utah, Salt Lake City, Utah, USA}
    \IEEEauthorblockA{\IEEEauthorrefmark{2}Idaho National Laboratory, Salt Lake City, Utah, USA\\}
\thanks{This manuscript has in part been authored by Battelle Energy Alliance, LLC under Contract No. DE-AC07-05ID14517 with the U.S. Department of Energy. The United States Government retains and the publisher, by accepting the paper for publication, acknowledges that the United States Government retains a nonexclusive, paid-up, irrevocable, world-wide license to publish or reproduce the published form of this manuscript, or allow others to do so, for United States Government purposes. \bf{STI Number: INL/CON-20-57149}.}
}

\maketitle
\begin{abstract}
Cyclic Prefix Direct Sequence Spread Spectrum (CP-DSSS) is a promising solution for futuristic 6G ultra-reliable low latency communications (URLLC) and massive machine type communication (mMTC) applications.  We propose that in such applications, the CP-DSSS waveform would operate as a secondary network at the same frequencies as the primary network but at much lower SNR.  In this paper, we evaluate per-user capacity of CP-DSSS when simple matched filtering (MF) is performed on the uplink (UL) and time-reversal (TR) precoding is used on the downlink (DL).  In this setting when operating in the low SNR regime, CP-DSSS achieves a per-user capacity that is near the optimum single-user capacity.  TR precoding converges to the optimal capacity as the number of antennas at the hub/gateway increases.  Using the estimated channel impulse response for MF and TR introduces little to no capacity loss.  Given the near-optimal performance of MF detection and TR precoding for each of the users, CP-DSSS can be implemented with simple device transceiver structures, reducing per-unit cost for massively deployed 6G networks.
\end{abstract}

\section{Introduction}
The cyclic prefix direct sequence spread spectrum (CP-DSSS) waveform is a promising solution for ultra-reliable low latency communications (URLLC) and massive machine type communications (mMTC), especially in the context of a secondary network waveform, operating at the same center frequency as the primary network.  In order to share the same spectrum, CP-DSSS would operate at a low signal to noise ratio (SNR) at reduced bit rates.  As described in \cite{Aminjavaheri:2018:CP_DSSS_Intro} and \cite{Farhang:2019:Scheduling_URLLC}, the cyclic prefix used in CP-DSSS allows a number of processing simplifications, which result from properties of circulant matrices.  In addition, further simplifications are afforded to user terminals when only the hub or gateway is responsible for using channel state information (CSI) to receive uplink (UL) signals or precode downlink (DL) signals.  This paper shows how matched filtering (MF) detection and time reversal (TR) precoding methods can be applied to a multi-user scenario to achieve near-optimal per-user capacity when operating at low SNRs, particularly when the hub or gateway is equipped with multiple antennas.

This paper shows the per-user capacity of MF detection for UL and TR precoding for DL averaged over a number of dispersive channels using ideal and realistic CSI.  In the low SNR regime (e.g., $<-20$ dB), the noise terms are dominant compared to interference from non-orthogonal detection.  This paper shows that interference contributions from additional users behave much like the non-orthogonal self-interference for a single user.  Consequently, for a reasonable number of active users (e.g., $K=32$), the per-user capacity results are extremely close to the single-user capacity results in the low SNR regime.  It follows that for low SNRs, the sum-capacity is approximately $K$ times larger than the single user capacity, where $K$ is the number of users.  For a given number of users, the signal to interference plus noise ratio (SINR) of each user is evaluated for a number of channel realizations, and the results are then converted to capacity.  These results are presented along with the ideal-receiver single-user capacity calculated in \cite{Kenney:2020:CP-DSSS_Capacity} for comparison.  CP-DSSS offers a very convenient method for channel estimation.  It is found that the robust behavior of CP-DSSS allows use of channel estimates obtained in the low operating SNR, instead of ideal CSI, with a negligible drop in performance.

The results presented in this paper show a simultaneous user load up to 32 users (i.e., $K=32$).  The data trends suggest that more simultaneous users would be possible.  As discussed in \cite{Kenney:2020:CP-DSSS_Capacity}, CP-DSSS is well-suited for a femtocell scenario, where the network density can be dramatically increased by supporting several user terminals in a small geographic area (e.g., $10$’s of meters), serviced by the femtocell gateway (FGW).  This allows for several FGW's to be distributed throughout the traditional cell.  The femtocells operate at relatively low power, thereby mitigating additional interference to the primary network.  One objective of the femtocell scenario is to minimize the complexity of the femtocell end terminals in order to reduce cost.  When MF detection and TR precoding are performed at the FGW, no CSI is needed at the femtocell terminal.  This results from the assumption of channel reciprocity in a time domain duplexing (TDD) scenario.  At the same time, both the FGW and the femtocell terminals must be able to coexist with the potentially strong signal levels from the primary network.  This situation can be addressed in at least two ways—either the CP-DSSS receivers treat the primary network signal as noise, or they detect the signal and subtract off its contribution to the overall received signal.  The former solution further decreases the SINR operating point, while the latter increases the complexity of the receiver.  The impact of these two potential solutions will be a subject of a future study.

Another point that should be noted here is the fact that in a FGW with multiple antennas, the MF detectors and TR precoders, as well as channel estimation, are set independently for each antenna. There is no joint detection or joint precoding.  This allows straightforward distribution of antennas within each femtocell for further improvement of the network, much in the same way as distributed antenna systems mentioned in \cite{Heath:2013:Distributed_Antennas} and \cite{Ngo:2017:Cell_Free}.

The rest of this paper is organized as follows.  Section II describes aspects of the CP-DSSS waveform that are of particular importance to this performance analysis.  Specifically, the methods of precoding, symbol rate reduction, the corresponding mathematical models for single antenna and multiple antenna operation, and channel estimation are discussed.  The ideal capacity of a single-user scenario with an ideal receiver is reviewed in Section III.  Section IV explains how the per-user capacity is calculated for a single-antenna FGW.  In Section V, the results are expanded to a multiple-antenna FGW scenario.  Section VI shows results with the exact users' channels replaced with their estimates.  Finally, Section VII provides concluding remarks.

\section{Waveform Summary}
The authors in \cite{Aminjavaheri:2018:CP_DSSS_Intro} provided details of the CP-DSSS waveform spreading and despreading.  A summary of the waveform behavior is presented here for completeness.  The spreading sequences used by CP-DSSS are from the family of Zadoff-Chu (ZC) sequences, which have the property of orthogonality between cyclically shifted versions of the same sequence.  Let the vector $\mathbf{z}_{(0)}$ represent a ZC sequence of length $N$ scaled to unit power (i.e., $\mathbf{z}_{(0)}^{\textrm{H}} \mathbf{z}_{(0)} = 1$), where the subscript references the size of the cyclic shift and superscript $\textrm{H}$ represents Hermetian.  Since each cyclic shift of $\mathbf{z}_{(0)}$ is orthogonal to the other $N-1$ cyclic shifts, there is a potential of modulating $N$ symbols per frame, one on each of the cyclically shifted ZC vectors.  We define the spreading matrix $\mathbf{Z}$ as
\begin{equation}
\mathbf{Z}=
\begingroup 
\setlength\arraycolsep{4pt}
\begin{bmatrix}
\mathbf{z}_{(0)} & \mathbf{z}_{(1)} & \dots & \mathbf{z}_{(N-2)} & \mathbf{z}_{(N-1)}
\end{bmatrix}
\endgroup. \label{eq:01}
\end{equation}

The corresponding despreading matrix is $\mathbf{Z}^{\rm{H}}$, noting that $\mathbf{Z}^{\textrm{H}}$ is the inverse of $\mathbf{Z}$.  Another important property of $\mathbf{Z}$ and $\mathbf{Z}^{\textrm{H}}$ is that they are circulant matrices; hence, they can be diagonalized by the Discrete Fourier Transform (DFT) matrix, represented as $\Fourier$, and the Inverse DFT (IDFT) matrix, represented as $\Fourier^{-1}$ (i.e., $\mathbf{Z}=\Fourier^{-1} \mathbf{\Lambda}_Z \Fourier$ and $\mathbf{Z}^{\textrm{H}}=\Fourier^{-1} \mathbf{\Lambda}_Z^{-1} \Fourier$, where $\mathbf{\Lambda}_Z$ is a diagonal matrix).  The real advantage of these circulant matrices for spreading and despreading is that they can be implemented in a computationally efficient means by using the Fast Fourier Transform (FFT) and Inverse Fourier Transform (IFFT), resulting in complexity of $\mathcal{O}(N \textrm{log} N)$ instead of $\mathcal{O}(N^2)$.

The transmitted signal is formed by premultiplying a vector $\mathbf{s}$ of $N$ data symbols by the spreading matrix and taking a duplicate of the last $N_{cp}$ samples to be transmitted first as a cyclic prefix (CP).  The length of $N_{cp}$ must be greater than or equal to the maximum delay spread of the channel in order to preserve the properties of circular convolution as in the case of OFDM in a 4G LTE or 5G NR context.  For the purposes of this paper, we assume that $N$ and $N_{cp}$ are identical to OFDM parameters, where $N$ is the number of OFDM subcarriers.  The signal seen by the receiver can now be represented as
\begin{equation}
\mathbf{y}=\mathbf{HZs}+\mathbf{v} \label{eq:02}
\end{equation}
after the CP has been removed, where $\mathbf{H}$ is the circulant channel matrix.  $\mathbf{H}$ is formed by taking the channel impulse response, $\mathbf{h}$, of length $L_h$, appending $N-L_h$ zeros to form $\mathbf{h}_{(0)}$, and then taking cyclic shifts of $\mathbf{h}_{(0)}$ to create
\begin{equation}
\mathbf{H}=
\begingroup 
\setlength\arraycolsep{4pt}
\begin{bmatrix}
\mathbf{h}_{(0)} & \mathbf{h}_{(1)} & \dots & \mathbf{h}_{(N-2)} & \mathbf{h}_{(N-1)}
\end{bmatrix}
\endgroup, \label{eq:03}
\end{equation}
as was done to form $\mathbf{Z}$ in \eqref{eq:01}.

To despread the received signal, $\mathbf{y}$ is left multiplied by $\mathbf{Z}^{\textrm{H}}$ to create $\tilde{\mathbf{y}}$.  Because $\mathbf{H}$ is circulant, it can also be diagonalized by the DFT matrix such that $\mathbf{Z}^{\textrm{H}} \mathbf{HZ} = ( \Fourier^{-1} \mathbf{\Lambda}_Z^{-1} \Fourier ) ( \Fourier^{-1} \mathbf{\Lambda}_H \Fourier ) ( \Fourier^{-1} \mathbf{\Lambda}_Z \Fourier ) = \Fourier^{-1} \mathbf{\Lambda}_Z^{-1} \mathbf{\Lambda}_H \mathbf{\Lambda}_Z \Fourier$.  Since diagonal matrices are commutable and $\mathbf{\Lambda}_Z^{-1} \mathbf{\Lambda}_Z = \mathbf{I}_N$, the despread signal can be represented as
\begin{equation}
\tilde{\mathbf{y}}=\mathbf{Hs}+\tilde{\mathbf{v}}, \label{eq:04}
\end{equation}
where $\tilde{\mathbf{v}}$ is the noise vector ($\sim \mathcal{N}(0,\sigma_{\tilde{v}}^2)$) after being left multiplied by $\mathbf{Z}^{\textrm H}$.  Since $\mathbf{Z}^{\textrm{H}}$ is a unitary matrix (i.e., complex orthogonal columns that are unit power), it does not change the noise statistics (i.e., $\sigma_{\tilde{v}}^2=\sigma_v^2$).

\subsection{Precoding}
CP-DSSS has the ability to scale the power on specific frequencies with the same fidelity as OFDM.  The power scaling is performed by multiplying the transmitted symbols by a precoding matrix, $\mathbf{G}$, prior to spreading.  The resulting received signal with precoding is
\begin{equation}
\tilde{\mathbf{y}}=\mathbf{HGs}+\tilde{\mathbf{v}}, \label{eq:05}
\end{equation}  
and the effective channel seen by the receiver is now $\mathbf{HG}$.  In order to keep the  transmitted power constant, there is a constraint that $\textrm{tr}(\mathbf{G}^{\textrm H}\mathbf{G})=N$.  Another constraint applied to facilitate analysis with precoding in \cite{Kenney:2020:CP-DSSS_Capacity} is to make $\mathbf{G}$ circulant.  It should be noted that this last constraint does not limit the precoder's ability to scale power on a frequency bin basis.

Two precoding options were discussed in \cite{Kenney:2020:CP-DSSS_Capacity}.  The first precoder used the water-filling (WF) result, which has been shown to be optimal in terms of capacity for OFDM.  It is instructive to note that WF emphasizes strong frequencies of the channel and does not transmit power at frequencies below a calculated threshold \cite{Goldsmith:2015}.  The second precoder used a simple TR filter, which was shown to significantly improve capacity over the equal power case (i.e., no precoding).  The TR precoder also emphasizes strong frequencies of the channel, but it de-emphasizes weaker frequencies of the channel instead of cutting them off as in WF.  The TR precoding technique will be examined in further detail in a later section.

\subsection{Symbol Rate Reduction} \label{sec:symbol_rate_reduction}
Although capacity is generally maximized when $N$ symbols are sent per CP-DSSS frame, there are advantages to reducing the symbol rate.  For example, if the capacity calculation shows that the number of bits per symbol is very low (e.g., leading to a coding rate of less than 0.1), then a FEC scheme of high complexity must be used to encode/decode the data, resulting in higher transceiver complexity.  Reducing the symbol rate allows more power to be transmitted per symbol, while still maintaining the same SNR.  Consequently, higher bit to symbol ratios can be used that are within the range of today's FEC schemes.  We may recall that coding rates in the range of 0.2 to 0.83 for 5G NR are common \cite{Dahlman:2018:5G}.  Another reason for reducing the symbol rate in CP-DSSS is to spread out the symbols, so that there is less impact from inter-symbol interference (ISI).

As described in \cite{Kenney:2020:CP-DSSS_Capacity}, symbol rate reduction is accomplished by forming an expander matrix, $\mathbf{E}_L$, which is based on the symbol reduction factor, $L$.  The form of $\mathbf{E}_L$ can be described as an identity matrix of dimension $N/L$ (i.e., $\mathbf{I}_{N/L}$) that has been upsampled in the vertical dimension by a factor of $L$.  In other words, after each row of $\mathbf{I}_{N/L}$, $L-1$ rows of zeros are inserted, resulting in an $N \times N/L$ matrix.  When symbol reduction and precoding are employed, the received signal takes the form
\begin{equation}
\tilde{\mathbf{y}}=\mathbf{HGE}_L \mathbf{s}+\tilde{\mathbf{v}}, \label{eq:06}
\end{equation}  
where $\mathbf{s}$ has $N/L$ symbols and each symbol is scaled by $\sqrt{L}$ such that the power per symbol is $L$ times $\sigma_s^2$.

\subsection{Multiple Antenna Scenario}
Operation with multiple antennas at the base station allows more simultaneous single antenna users to be supported and also reduces the transmitted power level.  The system model shown in \eqref{eq:06} can be modified with the following substitutions to facilitate multiple receive antennas in an UL scenario:
\begin{equation}
\tilde{\mathbf{y}}=
  \begin{bmatrix}
    \tilde{\mathbf{y}}^{(1)} \\
    \tilde{\mathbf{y}}^{(2)} \\
    \vdots \\
    \tilde{\mathbf{y}}^{(M)}
  \end{bmatrix}, \\
\mathbf{H}=
  \begin{bmatrix}
    \mathbf{H}^{(1)} \\
    \mathbf{H}^{(2)} \\
    \vdots \\
    \mathbf{H}^{(M)}
  \end{bmatrix}, \\
\tilde{\mathbf{v}}=
  \begin{bmatrix}
    \tilde{\mathbf{v}}^{(1)} \\
    \tilde{\mathbf{v}}^{(2)} \\
    \vdots \\
    \tilde{\mathbf{v}}^{(M)}
  \end{bmatrix},  
 \label{eq:07}
\end{equation}
where $M$ is the number of receive antennas, $\tilde{\mathbf{y}}^{(1)}$ through $\tilde{\mathbf{y}}^{(M)}$ are received signal vectors at the specified antennas, $\mathbf{H}^{(1)}$ through $\mathbf{H}^{(M)}$ are the circulant channel matrices corresponding to each antenna, and $\tilde{\mathbf{v}}^{(1)}$ through $\tilde{\mathbf{v}}^{(M)}$ are the noise vectors for each antenna.  No precoding is assumed for the UL scenario.

For the DL scenario, precoding is expected at the transmitter and the following substitutions are made into the system model in \eqref{eq:06}:
\begin{equation}
\mathbf{H} = 
	\begingroup 
		\setlength\arraycolsep{4pt}
		\begin{bmatrix}
		\mathbf{H}^{(1)} & \mathbf{H}^{(2)} & \dots & \mathbf{H}^{(M)}
		\end{bmatrix},
	\endgroup  
\mathbf{G}=
	\begin{bmatrix}
    	\mathbf{G}^{(1)} \\
	    \mathbf{G}^{(2)} \\
	    \vdots \\
	    \mathbf{G}^{(M)}
	\end{bmatrix}, \\
\label{eq:08}
\end{equation}
where the same conventions are used as in \eqref{eq:07}, and $\mathbf{G}^{(1)}$ through $\mathbf{G}^{(M)}$ are the precoding matrices corresponding to each antenna element.

\subsection{Channel Estimation} 
The objective of channel estimation is to estimate the channel impulse response, which can be a difficult task when operating at low SNRs.  By using the symbol rate reduction feature of CP-DSSS with $L=N$, a single symbol is sent in the pilot frame.  The same overall SNR is achieved by increasing the amplitude of the pilot symbol by a factor of $\sqrt{N}$, and the flat spectral response of the transmitted pilot is maintained through the ZC spreading.  The cyclic shift of the ZC spreading sequence is selected to be unique for each user transmitting a pilot sequence and should be sufficiently spaced to minimize overlap from different pilots.  Given an example with $K=32$ users and $N=2048$, the ZC delays should be spaced by $N/K=64$ samples.  In this example, there could still be some overlap between users depending on the delay spread of the channel impulse reponses, but the receiver uses the first 64 samples and the overlap, if any, adds to the estimation error.

On the receiver side, the channel impulse response is estimated after despreading the received signal and taking the resulting samples corresponding to each user's pilot signal.  The length of the estimated channel response is the minimum of $N/K$ and the CP length, $N_{\textrm{CP}}$.  As a final step, the channel is scaled by the inverse of the amplitude of the transmitted pilot symbol (i.e., $1/\sqrt{N}$).  As a result of being able to send a single symbol with the same combined power as a regular data frame, the effective SNR per sample is increased by as much as $10\textrm{log}_{10}(2048)=33.1$~dB.\label{chan_est}

\section{Capacity of an Ideal Receiver}
The capacity of an ideal CP-DSSS receiver is based on the mutual information of the received signal and the transmitted symbols without applying the limitations of a specific detector structure, and was derived in \cite{Kenney:2020:CP-DSSS_Capacity}.  Using the general system model \eqref{eq:06} with the results of \cite{Kenney:2020:CP-DSSS_Capacity} allows us to express the ideal capacity as
\begin{equation}
C=W \textrm{log}_2 \left( \lvert \mathbf{I}_{N} + L \frac{\sigma_s^2}{\sigma_{\tilde{v}}^2} \mathbf{HGE}_L\mathbf{E}_L^{\textrm{H}}\mathbf{G}^{\textrm{H}}\mathbf{H}^{\textrm{H}} \rvert \right) \label{eq:09}
\end{equation}
where $W$ is the bandwidth of the signal and $\mathbf{I}_N$ is the identity matrix of dimension $N$.

The ideal capacity is highest when $L=1$ as shown in \cite{Kenney:2020:MF_Detection}.  As $L$ increases, the rate at which the capacity rolls off depends on the SNR.  For low SNR cases (e.g., $-20$~dB), there is only a slight capacity reduction for moderate $L$ values (e.g., $L=32$).  All of the capacity results presented in this paper were simulated with $L=1$.  The results simulated in this paper use $N=2048$ with $130$ samples per channel, using an exponential power roll-off time constant of 25 samples (i.e., the power ratio between the first sample and the last is $\textrm{exp}((130-1)/25)\approx 174$ on average).  The relatively high exponential power roll-off time constant is used to show the robust nature of MF detection and TR precoding.  The curves reported in this paper represent average capacity taken over several randomly generated scenarios.  All users have the same channel gain, simulating the case where power control is employed.

\section{Single Antenna Multi-User Operation}
The first scenario to consider in a FGW or similar scenario is where the FGW has a single antenna.  It is assumed for this scenario and all others that each femtocell terminal also has a single antenna.  For both the MF detection and the TR precoding, the Hermetian of the channel matrix of a user $k$ is used to emphasize that user's signal and de-emphasize the signals from other users.  In the event that another user's channel is highly correlated with user $k$'s channel, then the interference from the highly correlated user will be more pronounced.  This scenario is taken into account by randomly selecting channels and averaging performance of many iterations.

In order to simulate the capacity for this multi-user scenario, we calculate the SINR over several channel instantiations.  This is achieved by subtracting the estimated symbols of user $k$, $\hat{\mathbf{s}}_k$, from the actual symbols from user $k$, $\mathbf{s}_k$.  The variance of the difference is the inverse of the SINR for user $k$, $\rho_k$.  The per-user capacity is then calculated using Shannon's capacity theorem,
\begin{equation}
C=\frac{WN}{L} \textrm{log}_2 \left( 1 + \rho_k \right), \label{eq:10a}
\end{equation}
where $W$ is the sample rate, $N$ is the number of payload samples in a frame, and $L$ is the rate reduction parameter (i.e., $\frac{N}{L}$ symbols are sent by user $k$ in each frame).  It is noted that the rate reduction parameter, $L$, also has an impact on $\rho_k$ since the signal power is $L$ times stronger per symbol.  The following subsections give details of the UL and DL cases for the single antenna, multi-user scenario. 

\subsection{Single Antenna UL}
When calculating the capacity for the UL case, no precoding is performed by the femtocell terminals because it is assumed that terminals do not have CSI.  The FGW performs MF detection using the CSI that it obtains during the channel estimation process, where the terminals transmit pilot signals.  Although the initial results assume perfect CSI at the FGW, we represent the estimated channel matrix from user $k$ to the FGW as $\hat{\mathbf{H}}_i$ in order to distinguish it from the actual channel matrix from user $k$ to the FGW, $\mathbf{H}_i$.  Leveraging the simple channel model expressed in \eqref{eq:04}, we express the single antenna, multi-user UL received signal as
\begin{equation}
\tilde{\mathbf{y}}=\sum^{K}_{k=1}\mathbf{H}_k \mathbf{s}_k+\tilde{\mathbf{v}}, \label{eq:10b}
\end{equation}
where $K$ is the number of users.  The corresponding symbol vector estimate using MF detection for user $i$ is given by
\begin{equation}
\hat{\mathbf{s}}_i = \frac{\hat{\mathbf{H}}_i^{\textrm{H}} \tilde{\mathbf{y}}}{\hat{\mathbf{h}}_i^{\textrm{H}}\hat{\mathbf{h}}_i} = \frac{\hat{\mathbf{H}}_i^{\textrm{H}}\mathbf{H}_i \mathbf{s}_i}{\hat{\mathbf{h}}_i^{\textrm{H}}\hat{\mathbf{h}}_i} + \sum_{k=1,k\neq i}^K \frac{ \hat{\mathbf{H}}_i^{\textrm{H}} \mathbf{H}_k \mathbf{s}_k}{\hat{\mathbf{h}}_i^{\textrm{H}}\hat{\mathbf{h}}_i} + \frac{\hat{\mathbf{H}}_i^{\textrm{H}}\tilde{\mathbf{v}}}{\hat{\mathbf{h}}_i^{\textrm{H}}\hat{\mathbf{h}}_i}, \label{eq:10c}
\end{equation}
where $\hat{\mathbf{h}}_i$ is the estimated channel impulse response for user $i$.

\begin{figure}[!t]
\centering
\includegraphics[width=3.6in, clip=true, trim=4cm 8.5cm 4cm 8.5cm ]{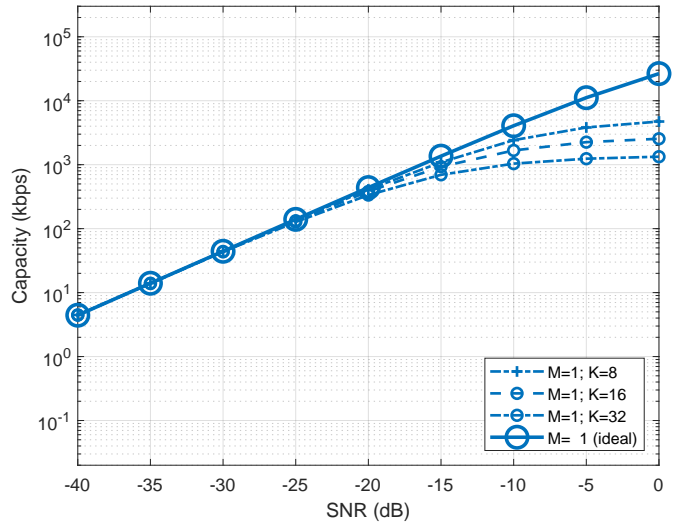}
\caption{UL per-user capacity for a single-antenna ($M=1$) FGW with an ideal detector (single user) and per-user capacity with MF detection ($K$ users) using ideal CSI.  Simulated results for the DL case are nearly identical to the UL, although the ideal detector capacity is higher for the DL due to precoding. }
\label{fig:single_antenna_UL}
\end{figure}

Fig. \ref{fig:single_antenna_UL} shows the per-user capacity for the UL case along with the ideal capacity for a single user.  Note that the simulated per-user capacity holds tight with the ideal single user capacity for all SNR values $\leq -25$ dB.  Even at an SNR of $\leq -20$ dB, the difference between the ideal single user detector and the 32 user case is very slight.

\subsection{Single Antenna DL}
Precoding is an essential part of preparing the DL signal for transmission because it is assumed that the femtocell terminals do not have CSI to use for detection.  The FGW is assumed to transmit user $k$'s DL signal with the same power that user $k$ transmitted the uplink signal.  As a result, the composite DL signal is $K$ times stronger than the UL signal from a single user in order to preserve link symmetry.  In practice, the total output power at the FGW can be reduced at the expense of SNR if the interference is deemed to be excessive, but the link symmetry assumption is instructive for this paper because it allows comparison between the UL and DL cases.  The signal received by user $i$ is given as
\begin{equation}
\tilde{\mathbf{y}}_i = \mathbf{H}_i \sum^{K}_{k=1} \hat{\mathbf{H}}^{\textrm{H}}_k \mathbf{s}_k + \tilde{\mathbf{v}}_i. \label{eq:10d}
\end{equation}

Unlike the UL case where the CSI is available at the receiver, the detected symbols are scaled by the estimate of the channel power arrived at by a gain control circuit or by other means.  For convenience, this scaling value is represented as $\hat{\mathbf{h}}_k^{\textrm{H}}\hat{\mathbf{h}}_k$ to be consistent with the UL case.  The corresponding symbol vector estimate is given by
\begin{equation}
\hat{\mathbf{s}}_i = \frac{\tilde{\mathbf{y}}_i}{\hat{\mathbf{h}}_i^{\textrm{H}}\hat{\mathbf{h}}_i} = \frac{\mathbf{H}_i \hat{\mathbf{H}}_i^{\textrm{H}} \mathbf{s}_i}{\hat{\mathbf{h}}_i^{\textrm{H}}\hat{\mathbf{h}}_i} +  \sum^{K}_{k=1, k\neq i} \frac{ \mathbf{H}_i \hat{\mathbf{H}}^{\textrm{H}}_k \mathbf{s}_k}{\hat{\mathbf{h}}_i^{\textrm{H}}\hat{\mathbf{h}}_i} + \frac{\tilde{\mathbf{v}}_i}{\hat{\mathbf{h}}_i^{\textrm{H}}\hat{\mathbf{h}}_i}. \label{eq:10e}
\end{equation}

The simulated DL capacity values are nearly  identical to the UL capacity values shown in Fig. \ref{fig:single_antenna_UL}.  However, there is a significant difference in the ideal capacity curve, since it assumes the benefit of precoding with an ideal detector.  If the femtocell terminal were to employ a MF detector using the composite channel (i.e., TR precoding convolved with the channel impulse response), then the simulated capacity would be close to the calculated ideal capacity.  Because no CSI is assumed at the receiver in the DL case, the expected disparity remains but will be reconciled when multiple FGW antennas are considered.


\section{Multiple Antennas Multi-User Operation}
In the next scenario the FGW is outfitted with multiple antennas and the femtocell terminals still have a single antenna.  When the number of FGW antennas, $M$, exceeds the number of simultaneous users, the convention used throughout this paper to have the femtocell terminals send the pilot signals proves to be more efficient than the opposite direction.  The same per-user capacity calculation used in \eqref{eq:10a} is also used for the multiple antenna case.

\subsection{Multiple Antenna UL}
As in the single antenna case, no precoding is performed by the femtocell terminals because it is assumed that the terminals do not have CSI.  The received signal at antenna $m$ is expressed as follows:
\begin{equation}
\tilde{\mathbf{y}}^{(m)} = \sum^{K}_{k=1} \mathbf{H}_k^{(m)} \mathbf{s}_k + \tilde{\mathbf{v}}^{(m)}, \label{eq:10f}
\end{equation}
where $\mathbf{H}_k^{(m)}$ is the channel matrix between user $k$ to antenna $m$ of the FGW and $\tilde{\mathbf{v}}^{(m)}$ is the noise vector at antenna $m$ after despreading.  The FGW performs MF detection at each antenna using the CSI that it obtains during the channel estimation process, which is represented in the form of the channel matrix, $\hat{\mathbf{H}}_k^{(m)}$ for user $k$ and antenna $m$.  The MF outputs from each antenna are then averaged to arrive at the symbol vector estimate, which is given as
\begin{equation}
\hat{\mathbf{s}}_i = 
\frac{1}{M} \sum_{m=1}^M \left( \frac{\hat{\mathbf{H}}_i^{(m)\textrm{H}}}{\hat{\mathbf{h}}_i^{(m)\textrm{H}}\hat{\mathbf{h}}_i^{(m)}} \left( \sum_{k=1}^K \mathbf{H}_k^{(m)} \mathbf{s}_k + \tilde{\mathbf{v}}^{(m)} \right) \right), \label{eq:10g}
\end{equation}
where $\hat{\mathbf{h}}_i^{(m)}$ is the estimated channel for user $i$ at antenna $m$.

The simulated per-user capacity for the UL case are nearly identical to the DL case shown in the next section for for varying values of $M$.  Similar to the single antenna case, the simulated per-user capacity holds tight with the ideal single-user capacity for all SNR values $\leq -25$ dB no matter the value of $M$.  Even at an SNR of $\leq -20$ dB, the difference between the ideal and 32 users is very slight.  Another promising result is that per-user capacity increases linearly with $M$ in the low SNR regime with only a slight degradation for the number of users.


\subsection{Multiple Antenna DL}
The precoding used for DL transmission is performed on a per-antenna basis.  The objective of TR precoding is to emphasize the signal at the receiver.  When this is conducted by multiple antennas at the FGW, the received signals from each antenna can be added constructively, providing the expected array gain from the transmitter.  In order to maintain the same transmit power, the signal power is divided equally between all transmitting antennas.  Signals destined for different users are multiplexed together at each antenna.  The signal received by user $i$ can be expressed as
\begin{equation}
\tilde{\mathbf{y}}_i = \sum^{M}_{m=1} \sum^{K}_{k=1} \frac{\mathbf{H}_i^{(m)} \hat{\mathbf{H}}^{(m)\textrm{H}}_k \mathbf{s}_k}{\sqrt{M}} + \tilde{\mathbf{v}}_i. \label{eq:10h}
\end{equation}

After the $M$ copies of the signal are received, they must be scaled by $1/\sqrt{M}$ along with the noise, which yields the $M$ factor increase in the SINR.  The channel scaling must also be accounted for as in the single antenna case.  The corresponding symbol vector estimate is given by
\begin{equation}
\hat{\mathbf{s}}_i = 
\sum_{m=1}^M \sum_{k=1}^K \left( \frac{\mathbf{H}_i^{(m)} \hat{\mathbf{H}}_k^{(m)\textrm{H}} \mathbf{s}_k}{M \hat{\mathbf{h}}_i^{(m)\textrm{H}}\hat{\mathbf{h}}_i^{(m)}} \right)+ \frac{\tilde{\mathbf{v}}_i^{(m)}}{\sqrt{M} \hat{\mathbf{h}}_i^{(m)\textrm{H}}\hat{\mathbf{h}}_i^{(m)}}, \label{eq:10i}
\end{equation}
where $\hat{\mathbf{h}}_i$ is the estimated channel for user $i$.

\begin{figure}[!t]
\centering
\includegraphics[width=3.5in, clip=true, trim=4cm 8.5cm 4.5cm 9.2cm ]{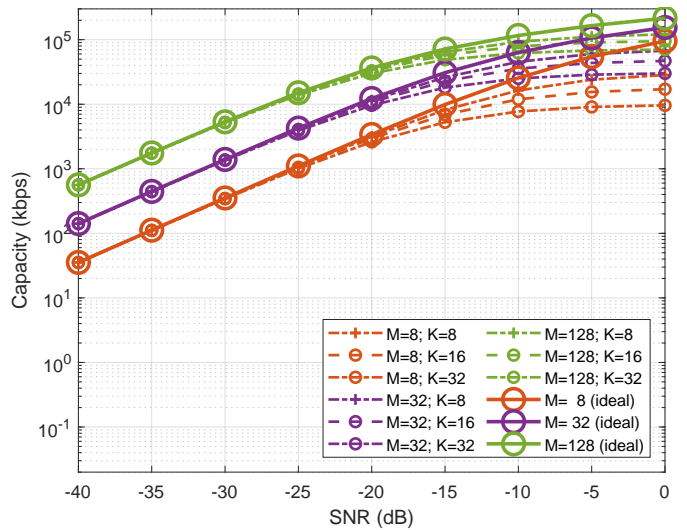}
\caption{DL per-user capacity of a multiple antenna FGW with an ideal detector (single user) and per-user capacity with MF detection ($K$ users).  Note that there is negligible difference between the multi-user curves in the low SNR regime. }
\label{fig:multiple_antenna_DL}
\end{figure}

The simulated per-user capacity values for the DL are shown in Fig. \ref{fig:multiple_antenna_DL}.  One interesting result with multiple antennas is that the precoding advantage of the ideal receiver over the detected per-user capacity diminishes as the number of FGW antennas increases.  For example, Fig. \ref{fig:multiple_antenna_DL} shows that the per-user capacity curves are basically the same as the ideal single user capacity curves for $M=32$ and above in the low SNR regime.

\section{Channel Estimation Impact}
\begin{figure}[!t]
\centering
\includegraphics[width=3.5in, clip=true, trim=4cm 8.5cm 4.5cm 9cm ]{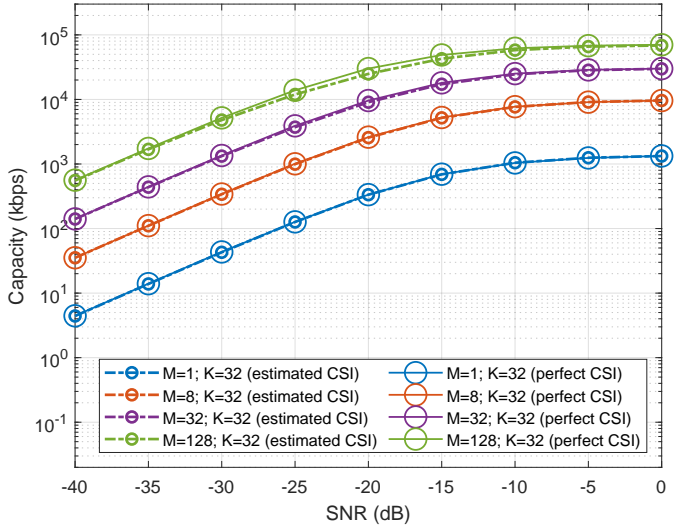}
\caption{Comparison of DL per-user capacity with and without channel estimation for the 32 user case.  Note that there is a slight degradation due to channel estimation for 128 antenna for SNR values between $-25$ and $-15$~dB, but lower antenna counts are negligibly affected.}
\label{fig:multiple_antenna_DL_w_chan_est}
\end{figure}

When the simple channel estimation shown in Section \ref{chan_est} is used instead of perfect CSI, the performance is largely unchanged over all SNR values of interest.  Fig. \ref{fig:multiple_antenna_DL_w_chan_est} compares the per-user channel capacity using TR precoding for the DL with estimated CSI and perfect CSI.  For lower antenna counts (e.g. $M=1$ or $M=8$) there is no measureable difference in capacity.  However, as the antenna count grows to $32$, some small differentiation is noticeable.  At $M=128$ the degradation due to the estimated CSI is more pronounced, resulting in a capacity reduction of about $18\%$ at $-20$~dB SNR.  This behavior can be explained by considering that the quality of the estimated channel is the same for each antenna at a given SNR.  At the same time, the array gain of the FGW has the potential to increase the SNR by $10 \textrm{log}_{10} M$, but it does not attain all of the gain because the array gain out-paces the quality of the channel estimate.  As the SNR increases, multi-user interference begins to dominate and the performance with estimated CSI converges again with the perfect CSI case.  Despite the small capacity reduction for $M=128$, these results show that the simple channel estimation technique used by CP-DSSS should be sufficient for the data channel in future URLLC and mMTC systems.

\section{Conclusion}
CP-DSSS has tremendous potential as a URLLC or mMTC solution where it would act as a secondary network, using the same spectrum as the primary network but operating at low SNR values in order to minimize interference with the primary network.  This paper extended the single user capacity analysis presented in \cite{Kenney:2020:CP-DSSS_Capacity} by showing per-user capacity for multi-user scenarios, using simple MF detection for UL and TR precoding for the DL.  Simulation results showed that per-user capacity is extremely close to the ideal single user capacity for SNR values $\leq -25$~dB, even with 32 simultaneous users.  This favorable result means that each user can attain $100$ kbps to $10$ Mbps data rates, depending on the number of antennas deployed at the FGW, which should meet the needs of future 6G URLLC and mMTC scenarios.  We found that with the simple channel estimation procedure outlined for CP-DSSS, essentially the same capacity results were obtained when compared to perfect CSI.  These impressive results with simple detection or precoding showed that CP-DSSS has great potential to provide 6G URLLC and mMTC solutions for many simultaneous users with simplified transceiver terminals.


\end{document}